\documentclass[aps,prb,showpacs,superscriptaddress,twocolumn]{revtex4}
\usepackage{amsmath}
\usepackage{bm}
\usepackage{bbm}
\begin{document}
\title{On the conservation of spin currents in spin-orbit coupled systems}
\author{R. Shen}
\affiliation{Department of Physics, The University of Hong Kong,
Pokfulam Road, Hong Kong, China} \affiliation{National Laboratory of
Solid State Microstructures and Department of Physics, Nanjing
University, Nanjing, 210093 China}
\author{Yan Chen}
\affiliation{Department of Physics, The University of Hong Kong,
Pokfulam Road, Hong Kong, China}
\author{Z. D. Wang}
\affiliation{Department of Physics, The University of Hong Kong,
Pokfulam Road, Hong Kong, China}
\affiliation{National Laboratory of
Solid State Microstructures and Department of Physics, Nanjing
University, Nanjing, 210093 China}
\begin{abstract}
Applying the Gordon-decomposition-like technique, the convective
spin current (CSC) is extracted from the total angular-momentum
current. The CSC describes the transport properties of the electron
spin and is conserved in the relativistic quantum mechanics approach
where the spin-orbit coupling has been intrinsically taken into
account. Arrestingly, in the presence of external electromagnetic
field, the component of the convective spin along the field remain
still conserved.  This conserved CSC is also derived for the first
time in the nonrelativistic limit using the Foldy-Wouthuysen
transformation.
\end{abstract}
\maketitle
Spintronics is a fast developing field in condensed matter
physics, where the research on the spin transport and manipulation
in solid state systems appears to be a central
subject.\cite{Wolf,Zutic} Typically, the spin transport is
strongly affected by a coupling of spin and orbital degrees of
freedom in semiconductors. Of particular interest is the spin Hall
effect in a spin-orbit coupled system. The spin Hall effect means
that a transverse spin current can be induced by applying an
electric field. Earlier studies focused on the extrinsic effects
due to the spin dependent scattering.\cite{Dyakonov,Hirsch,Zhang}
Recently, an intrinsic spin Hall effect (ISHE), which is resulted
from the spin-orbit coupled bands and may exist in the clean
systems, has attracted considerable
attentions.\cite{Murakami,Sinova,Rashba,Hu,Schliemann,Sinitsyn,
Bernevig,ShenSQ,Guo,Inoue,Mishchenko,Dimitrova,Chalaev,
Murakami3,Bernevig2,Sugimoto,Zhang2,Culcer,Murakami2,Sun,
Jin,ZhangP,Wang,WangY,Kato,Wunderlich} On the theoretical side,
the ISHE has been predicted in the \textit{p}-doped semiconductors
described by the Luttinger-like model\cite{Murakami} and in the
two-dimensional electron gas described by the Rashba-like
model\cite{Sinova}. The ISHE in the two-dimensional hole
gas\cite{Bernevig2}
and the impurity scattering effect on the ISHE
\cite{Inoue,Mishchenko,Dimitrova,
Chalaev,Murakami3,Bernevig2,Sugimoto} have also been analyzed.
 On the experimental side,
the spin accumulation related to the spin Hall effect has been
observed in the \textit{n}-type bulk semiconductor\cite{Kato} and
in the two-dimensional hole gas.\cite{Wunderlich}

The spin current is an important issue in discussing the spin Hall
effect and other spin transport phenomena. In spite of the
extensive investigation on the ISHE, there is still a fundamental
problem that what is the proper definition of the spin current
operator in the spin-orbit coupled system. The conventional
definition of the spin current operator is the anticommutator of
the Pauli spin matrix and the velocity of the
electron.\cite{Murakami,Sinova} The electron spin is not a
conserved quantity in the spin-orbit coupled system. As a result,
there is a source term, namely spin torque, in the continuity-like
equation related to the conventional spin
current.\cite{Sun,Jin,ZhangP} The spin torque represents the spin
precession caused by the external field and the spin-orbit
coupling. The presence of the nonzero source term in the
continuity-like equation makes the conventional spin current an
ill-defined quantity to describe the transport of spin. On the
other hand, the non-conserved part of the conventional spin
current has very fast dynamics and can not contribute to the low
frequency response.\cite{Murakami2} Therefore, it is essential to
define a conserved effective spin current from a first principle
for studying the spin transport properties in the spin-orbit
coupled system.

There are already some efforts focusing on the understanding of
the nonconservation of the conventional spin
current.\cite{Zhang2,Culcer,Murakami2,Sun,Jin,ZhangP,Wang,WangY}
In the \textit{p}-doped semiconductor described by the
Luttinger-like model, Murakami \textit{et al}\cite{Murakami2} have
proposed a conserved effective spin current of the spin $3/2$ hole
by modifying both the spin density and the spin current operator.
By rewriting the spin torque as the divergence of a torque dipole
density, another approach to the conserved effective spin current
was also proposed\cite{Culcer,ZhangP} where the spin density
operator remains unchanged and only the spin current operator is
modified. In this paper, motivated by the fact that the spin-orbit
coupling in the semiconductors comes from the relativistic
correction of the electron motion, we develop a conserved
effective spin current in the relativistic quantum mechanics
approach where the spin-orbit coupling has been intrinsically
taken into account. For a relativistic electron, the spin is not
conserved, but the total angular-momentum (TAM) consisting of both
spin and orbit angular-momentum is a conserved quantity in the
absence of  external fields. Therefore, there is the conserved TAM
current according to the Noether's theorem. Under the
Gordon-decomposition-like scheme, the TAM current can be divided
into two parts: the convective TAM current and the internal one.
It is found that the spin part of the convective TAM current,
namely the convective spin current (CSC) satisfies the continuity
equation. The CSC is not only conserved but also closely related
to the spin flux of the electron moving in the spatial space and
therefore is an appropriate candidate for the spin current
operator in discussing the spin transport properties. In the
presence of external electromagnetic field, neither the TAM nor
the CSC is conserved in a general case because of the mutual
exchange of the angular-momentum between the electron and the
external field. However, more intriguingly, we find that if both
the electric field and the magnetic field are along the same
direction, namely the $z$-direction, the $z$-components of the CSC
and the TAM still remain conserved. Remarkably, this is just the
case in the Rashba-like spin-orbit coupled system. As a result,
the CSC is also likely robust subject to the external field in
some sense. Finally,  by using the Foldy-Wouthuysen
transformation, the conserved CSC is derived in the
nonrelativistic (NR) limit, which has  a potential application in
spintronics.

We start with the brief review on the Gordon decomposition of the
probability current of the relativistic electron. For the electron
of the mass $m$ traveling in the three-dimensional free space, the
four-component wave function of the electron $\Psi$ satisfies the
Dirac equation
\begin{equation}\label{Dirac}
\textrm{i}\hbar\frac{\partial\Psi}{\partial
t}=\left(c\bm{\alpha}\bullet\bm{p}+mc^{2}\beta\right)\Psi ,
\end{equation}
where the momentum $\bm{p}=-\textrm{i}\hbar\bm{\nabla}$, the
matrices $\bm{\alpha}=\left(
\begin{smallmatrix}
0&\bm{\sigma}\\\bm{\sigma}&0
\end{smallmatrix}\right)$,
$\beta=\left(
\begin{smallmatrix}\mathbbmss{1}&0\\0&-\mathbbmss{1}
\end{smallmatrix}\right)$, $\bm{\sigma}$ is the Pauli spin matrix
and $\mathbbmss{1}$ is the $2\times 2$ unity matrix. The invariance
of the Dirac Eq. (\ref{Dirac}) under the U(1) gauge transformation
results in the continuity equation for the probability current
\begin{equation}\label{current}
\frac{\partial\rho}{\partial t}+\bm{\nabla}\bullet\bm{j}=0,
\end{equation}
where the probability density $\rho =\Psi^{\dag}\Psi$ and the
current $\bm{j}=c\Psi^{\dag}\bm{\alpha}\Psi$. The Dirac Eq.
(\ref{Dirac}) can be rewritten as
\begin{equation}\label{var_Dirac}
\Psi=\frac{1}{mc^{2}}\beta\left(\textrm{i}\hbar\frac{\partial}
{\partial t}-c\bm{\alpha}\bullet
\frac{\hbar}{\textrm{i}}\bm{\nabla}\right)\Psi .
\end{equation}
Substituting Eq. (\ref{var_Dirac}) into the definitions of $\rho$
and $\bm{j}$, one finds that the density and current can be divided
into the convective and internal parts, known as the Gordon
decomposition, \cite{Gordon,Baym} in the form
\begin{equation}\label{Gordon_decomposition}
\rho =\rho _{\textrm{c}}+\rho _{\textrm{i}}, \quad \bm{j} =\bm{j}
_{\textrm{c}}+\bm{j} _{\textrm{i}},
\end{equation}
\begin{equation}\label{convective}
\begin{split}
&\rho
_{\textrm{c}}=\frac{\text{i}\hbar}{2mc^{2}}\left(\overline{\Psi}
\frac{\partial\Psi}{\partial t}-\frac{\partial
\overline{\Psi}}{\partial t}\Psi\right),
\\
&\bm{j} _{\textrm{c}}=\frac{\hbar}{2m\text{i}}\left[\overline{\Psi}
\left(\bm{\nabla}\Psi\right)-
\left(\bm{\nabla}\overline{\Psi}\right)\Psi\right],
\end{split}
\end{equation}
\begin{equation}\label{internal}
\rho _{\textrm{i}}=-\bm{\nabla}\bullet \bm{P}, \quad \bm{j}
_{\textrm{i}}=c\bm{\nabla}\times \bm{M}+\frac{\partial
\bm{P}}{\partial t},
\end{equation}
where
\begin{equation}\label{MP}
\bm{M}=\frac{\hbar}{2mc}\overline{\Psi}\bm{\Sigma}\Psi, \quad
\bm{P}=\frac{\hbar}{2mc}\overline{\Psi}
\left(-\text{i}\bm{\alpha}\right)\Psi,
\end{equation}
and the matrix $\bm{\Sigma}=\left(
\begin{smallmatrix}
\bm{\sigma} & 0 \\
0 & \bm{\sigma}
\end{smallmatrix}\right)$ as well as
$\overline{\Psi}=\Psi ^{\dag}\beta$. The internal parts of the
density and current are closely related to the internal state of the
electron. For the electron of charge $q$, $q\bm{M}$ is essentially
the magnetic moment density of the electron arising from its spin
and $q\bm{P}$ is the electric polarization belonging to the spin.
The convective parts of the density and current are determined by
the rate of change of $\Psi$ in space and time and closely related
to the motion of the electron in the spatial space. Notice that the
convective and internal parts of the density are separately
conserved. For the NR limit, $\rho _{\textrm{i}}$ is zero up to the
zeroth order in ($1/c$), $\bm{j} _{\textrm{i}}$ is only a curl term
and has nothing to do with the electron accumulation, and the
convective parts of the current and the density are reduced to the
usual form in the NR quantum mechanics,
\begin{equation}\label{convective_nonrel}
\rho _{\textrm{c,NR}}=\psi^{\dag}\psi, \; \bm{j}
_{\textrm{c,NR}}=\frac{\hbar}{2m\text{i}}\left[\psi^{\dag}
\left(\bm{\nabla}\psi\right)-
\left(\bm{\nabla}\psi^{\dag}\right)\psi\right],
\end{equation}
with $\psi$ the two-component wave function of the electron.

Next, we turn to the spin current. According to the Noether's
theorem, the invariance of the Dirac Eq. (\ref{Dirac}) under the
Lorentz rotation leads to the continuity equation for the TAM
current\cite{Schwabl}
\begin{equation}\label{angular_current}
\frac{\partial\rho^{\gamma}}{\partial
t}+\bm{\nabla}\bullet\bm{j}^{\gamma}=0,
\end{equation}
where the density of the $\gamma$-component of the TAM is
\begin{equation}\label{density_angular_current}
\rho^{\gamma}=\frac{1}{2}\left\{\Psi^{\dag}
\mathcal{J}_{\gamma}\Psi+h.c.\right\},
\end{equation}
and the current of the $\gamma$-component of the TAM is
\begin{equation}\label{current_angular_current}
\bm{j}^{\gamma}=\frac{1}{2}c\left\{\Psi^{\dag}
\bm{\alpha}\mathcal{J}_{\gamma}\Psi+h.c.\right\},
\end{equation}
where the $\gamma$ component of the TAM operator
$\mathcal{J}_{\gamma}=((\hbar/2)\bm{\Sigma}+
\bm{r}\times\bm{p})_{\gamma}$ with the position operator $\bm{r}$
and the index $\gamma=x,y,z$. The TAM contains two parts, the spin
part $(\hbar/2)\bm{\Sigma}$ and the orbit part $\bm{r}\times\bm{p}$.
For a relativistic electron, unlike the TAM, neither the spin
density nor the orbit angular-momentum density is conserved alone.
Similar to the probability current, the TAM current can also be
divided into the convective and internal parts \cite{Hehl}
\begin{equation}\label{convective_angular}
\begin{split}
\rho^{\gamma} _{\textrm{c}}=&\frac{1}{2}\left[
\frac{\textrm{i}\hbar}{2mc^{2}}\left(\overline{\Psi}
\mathcal{J}_{\gamma}\frac{\partial\Psi}{\partial t}-\frac{\partial
\overline{\Psi}}{\partial t} \mathcal{J}_{\gamma}\Psi
\right)+h.c.\right],\\
\bm{j}^{\gamma} _{\textrm{c}}=&\frac{1}{2}\left\{
\frac{\hbar}{2m\textrm{i}}\left[\overline{\Psi}
\bm{\nabla}\left(\mathcal{J}_{\gamma}\Psi\right)
-\left(\bm{\nabla}\overline{\Psi}\right)\mathcal{J}_{\gamma}\Psi
\right]+h.c.\right\},
\end{split}
\end{equation}
\begin{equation}\label{internal_angular}
\rho ^{\gamma}_{\textrm{i}}=-\bm{\nabla}\bullet \bm{P}^{\gamma},
\quad \bm{j} ^{\gamma}_{\textrm{i}}=c\bm{\nabla}\times
\bm{M}^{\gamma}+\frac{\partial\bm{P}^{\gamma}}{\partial t},
\end{equation}
where
\begin{equation}\label{MP_angular}
\begin{split}
&\bm{M}^{\gamma}=\frac{1}{2}\left(\frac{\hbar}{2mc}
\overline{\Psi}\bm{\Sigma}\mathcal{J}_{\gamma}\Psi+h.c.\right),
\\
&\bm{P}^{\gamma}=\frac{1}{2}\left[\frac{\hbar}{2mc}\overline{\Psi}
\left(-\text{i}\bm{\alpha}\right)\mathcal{J}_{\gamma}\Psi
+h.c.\right].
\end{split}
\end{equation}
The convective and internal parts of the TAM current have the same
structure as those of the probability current. For the NR limit,
$\rho^{\gamma} _{\textrm{i}}$ is zero up to the zeroth order in
$(1/c)$, $\bm{j}^{\gamma} _{\textrm{i}}$ is only a curl term and has
nothing to do with the accumulation of the TAM. Let's consider the
spin part of the convective TAM current, namely the CSC, with the
density for the spin-$\gamma$ component
\begin{equation}\label{csd}
\rho^{\textrm{s-}\gamma}_{\textrm{c}}=
\frac{\textrm{i}\hbar^{2}}{4mc^{2}}\overline{\Psi}
\Sigma_{\gamma}\frac{\partial\Psi}{\partial t}+h.c.,
\end{equation}
and the current
\begin{equation}\label{csc}
\bm{j}^{\textrm{s-}\gamma} _{\textrm{c}}=
\frac{\hbar}{2m\textrm{i}}\left[\overline{\Psi}
\frac{\hbar}{2}\Sigma_{\gamma}\left(\bm{\nabla}\Psi\right)
-\left(\bm{\nabla}\overline{\Psi}\right)\frac{\hbar}{2}
\Sigma_{\gamma}\Psi \right].
\end{equation}
The CSC is determined by the rate of change of $\Psi$ in space and
time and ought to be closely related to the spin flux of the
electron moving in the spatial space. Moreover, noting that the
solution of the Dirac Eq. (\ref{Dirac}) is also a solution of the
Klein-Gordon equation, one can easily find that the CSC satisfies
the continuity equation
\begin{equation}\label{cs_continue}
\frac{\partial\rho^{\textrm{s-}\gamma}_{\textrm{c}}}{\partial
t}+\bm{\nabla}\bullet\bm{j}^{\textrm{s-}\gamma}_{\textrm{c}}=0,
\end{equation}
despite that the bare spin is not a conserved quantity for the
relativistic electron. Therefore, we argue that the CSC is just the
correct effective spin current for discussing the spin transport
properties of the relativistic electron.

The results we present above are valid for the electrons moving in
the free space. The influence of the external electromagnetic field
can be taken into account by making the transformation
\begin{equation}\label{pA}
\bm{p}\rightarrow\bm{\pi}=\bm{p}-\frac{q}{c}\bm{A},\quad
\textrm{i}\hbar\frac{\partial}{\partial t}\rightarrow
\textrm{i}\hbar\frac{\partial}{\partial t}-qV
\end{equation}
with the charge of the electron $q$, the scalar potential $V$ and
the vector potential $\bm{A}$. The density and current of the
convective spin in the external electromagnetic field are written as
\begin{equation}\label{csd_EM}
\rho^{\textrm{s-}\gamma}_{\textrm{c}}=\left(
\frac{\textrm{i}\hbar^{2}}{4mc^{2}}\overline{\Psi}
\Sigma_{\gamma}\frac{\partial\Psi}{\partial
t}+h.c.\right)-\frac{\hbar}{2mc^{2}}qV\overline{\Psi}
\Sigma_{\gamma}\Psi ,
\end{equation}
\begin{equation}\label{csc_EM}
\bm{j}^{\textrm{s-}\gamma} _{\textrm{c}}=
\frac{\hbar^{2}}{4m\textrm{i}}\left[\overline{\Psi}
\Sigma_{\gamma}\left(\bm{\nabla}\Psi\right)-h.c.\right]
-\frac{q\hbar\bm{A}}{2mc}\overline{\Psi}\Sigma_{\gamma}\Psi .
\end{equation}
In the external electromagnetic field, neither the TAM nor the CSC
is conserved generally because of the mutual exchange of the
angular-momentum between the electron and the external field. The
continuity-like equation of the CSC reads
\begin{equation}\label{cs_continue_EM}
\frac{\partial\rho^{\textrm{s-}\gamma}_{\textrm{c}}}{\partial
t}+\bm{\nabla}\bullet\bm{j}^{\textrm{s-}\gamma}_{\textrm{c}}
=\mu_{\textrm{B}}\overline{\Psi}\left(\bm{\Sigma}\times\bm{B}
-\textrm{i}\bm{\alpha}\times\bm{E}\right)_{\gamma}\Psi ,
\end{equation}
where the Bohr Magneton $\mu_{\textrm{B}}=q\hbar/2mc$, the magnetic
field $\bm{B}=\bm{\nabla}\times\bm{A}$, and the electric field
$\bm{E}=-(\partial \bm{A}/\partial t)-\bm{\nabla}V$. Although the
CSC is not conserved, the source and drain of the CSC in Eq.
(\ref{cs_continue_EM}) are only the external fields, while those of
the bare spin include the contributions from both the external
fields and the orbit angular-momentum. Therefore, the CSC is still
more suitable than the bare spin current for discussing the spin
transport of the relativistic electron in the external fields. For a
particular case that both the electric and magnetic fields are along
the same direction, namely the $z$ direction, one can find from Eq.
(\ref{cs_continue_EM}) that the $z$-component of the convective spin
is still conserved. This particular case is relevant to the
two-dimensional electron gas with the Rashba interaction where the
electron is free in the $x$-$y$ plane and the confinement potential
is only along the $z$ direction. The $z$-component of the spin is
also of the most interest in the Rashba system. In the following
text we are restricted to this particular case, where the CSC is
still conserved and can be served as the proper quantity for
discussing the spin transport of the relativistic electron.

In order to apply the idea of the CSC into the spintronics, the NR
limit of the CSC operator is needed. The Hamiltonian of the
relativistic electron in the external electromagnetic field reads
\begin{equation}\label{Hamiltonian}
H=c\bm{\alpha}\bullet\bm{\pi}+qV+mc^{2}\beta .
\end{equation}
The NR limit of the Hamiltonian (\ref{Hamiltonian}) can be made by
the Foldy-Wouthuysen transformation.\cite{Foldy, Schwabl} Up to the
$(1/c^{2})$ order, the NR Hamiltonian of the electron is given as
\begin{equation}\label{nonrel_H}
\begin{split}
H_{\textrm{NR}}=&\frac{1}{2m}\bm{\pi}^{2}+qV-
\mu_{\textrm{B}}\bm{\sigma}\bullet\bm{B}-\frac{q\hbar ^{2}}
{8m^{2}c^{2}}\bm{\nabla}\bullet\bm{E}\\
&-\frac{q\hbar}{8m^{2}c^{2}}\bm{\sigma}\bullet\left(\bm{E}\times
\bm{\pi}-\bm{\pi}\times\bm{E}\right)\\
&-\frac{1}{8m^{3}c^{2}}\bm{\pi}^{4}+
\frac{\mu_{\textrm{B}}^{2}}{2mc^{2}}
\left(\bm{E}^{2}-\bm{B}^{2}\right)\\
&+\frac{\mu_{\textrm{B}}}{4m^{2}c^{2}}
\left[\left(\bm{\sigma}\bullet\bm{B}\right)\bm{\pi}^{2}+
\bm{\pi}^{2}\left(\bm{\sigma}\bullet\bm{B}\right)\right],
\end{split}
\end{equation}
and the two-component wave function $\psi$, \textit{i.e.}, the large
component of the four-component spinor $\Psi$ in the
Foldy-Wouthuysen representation corresponding the positive energy
solutions, satisfies the Schr\"{o}dinger equation
$\textrm{i}\hbar(\partial\psi/\partial t) =H_{\textrm{NR}}\psi$. The
Schr\"{o}dinger equation and the Hamiltonian (\ref{nonrel_H}) are
the starting point for discussing the motion of the election in the
NR quantum mechanics including some relativistic corrections up to
the $(1/c^{2})$ order. We recall the physical meanings of each term
in Eq. (\ref{nonrel_H}). The first three terms are the NR kinetic
energy, the potential energy, and the Zeeman energy arising from the
electron spin, respectively. The fourth and fifth terms are the
so-called Darwin term and the spin-orbit interaction, respectively.
All other terms are of even higher orders.

With the help of the Hamiltonian (\ref{Hamiltonian}) and the Dirac
equation, the convective spin density in Eq. (\ref{csd_EM}) can be
rewritten as $\rho^{\textrm{s-}z}_{\textrm{c}}
=\left(\Psi^{\dag}\hat{\rho}^{\textrm{s-}z}_{\textrm{c}}\Psi
+h.c.\right)/2$, with the operator
\begin{equation}\label{csd_op}
\hat{\rho}^{\textrm{s-}z}_{\textrm{c}}=\frac{\hbar}{2}\Sigma_{z}-
\frac{\textrm{i}\hbar}{2mc}\beta\left(\bm{\alpha}\times\bm{\pi}
\right)_{z}.
\end{equation}
One can easily verify that $\hat{\rho}^{\textrm{s-}z}_{\textrm{c}}$
is an Hermitian operator and commutes with the Hamiltonian
(\ref{Hamiltonian}) and therefore is a conserved quantity. The
difference between the convective spin and the bare spin is the
second term in Eq. (\ref{csd_op}) which is a higher order
relativistic correction term coupling the large and small components
of the four-component spinor $\Psi$. The NR limit of the convective
spin density operator up to the $(1/c^{2})$ order, the same order as
that of the Hamiltonian, is obtained by the same unitary
transformations applied to the Hamiltonian (\ref{Hamiltonian}) in
the Foldy-Wouthuysen scheme. After the Foldy-Wouthuysen
transformation, the convective spin density is also reduced to an
even operator as the Hamiltonian does, which does not couple the
large and small components of $\Psi$. The NR convective spin density
operator acting on the two-component spinor $\psi$ is obtained as
\begin{equation}\label{cs_nr_op}
\begin{split}
\hat{\rho}^{\textrm{s-}z}_{\textrm{c,NR}}
&=\frac{\hbar}{2}\sigma_{z}+\frac{\textrm{i}\hbar}
{4m^{2}c^{2}}\left(\bm{\sigma}
\bullet\bm{\pi}\right)\left(\bm{\sigma}\times\bm{\pi}\right)_{z} \\
&=\frac{\hbar}{2}\sigma_{z}
+\frac{\hbar}{4m^{2}c^{2}}\hat{\mathcal{C}},
\end{split}
\end{equation}
with the operator
$\hat{\mathcal{C}}=\left(\pi_{x}^{2}+\pi_{y}^{2}\right)\sigma_{z}
-\left(\sigma_{x}\pi_{x}+\sigma_{y}\pi_{y}\right)\pi_{z}$. With the
help of the operator (\ref{cs_nr_op}) the convective spin density is
written as $\rho^{\textrm{s-}z}_{\textrm{c,NR}}
=\left(\psi^{\dag}\hat{\rho}^{\textrm{s-}z}_{\textrm{c,NR}}\psi
+h.c.\right)/2$. Due to the unitarity of the transformation, the NR
convective spin density operator Eq. (\ref{cs_nr_op}) commutes with
the Hamiltonian (\ref{nonrel_H}) and remains as the conserved
quantity up to the $(1/c^{2})$ order. The difference between the
convective spin and the bare spin is the second term in Eq.
(\ref{cs_nr_op}) which is of the order of $(1/c^{2})$ and is very
small. However, the spin-orbit interaction is also originated from
the relativistic correction of the electron motion up to the
$(1/c^{2})$ order. Therefore, the second term in Eq.
(\ref{cs_nr_op}) can not be neglected when discussing the spin
current in the spin-orbit coupled system.

Although there are many terms in the NR Hamiltonian
(\ref{nonrel_H}), only the spin-orbit term is of particular interest
on the topic of the spin Hall effect. Therefore, we remove the other
relativistic corrections and rewrite the effective Hamiltonian in
the form
\begin{equation}\label{H_spin-orbital}
H_{\textrm{SO}}=\frac{1}{2m}\bm{\pi}^{2}+qV-\mu_{\textrm{B}}
B\sigma_{z}+\frac{q\hbar E}{4m^{2}c^{2}}\left(\bm{\sigma}\times
\bm{\pi}\right)_{z}.
\end{equation}
It should also be noted that, in going from Eq. (\ref{nonrel_H}) to
Eq. (\ref{H_spin-orbital}), it has been assumed that the electric
field is static, \textit{i.e.}, $\bm{E}=-\bm{\nabla}V$, and the
external fields are along the $z$-direction. It can be easily
verified that $\hat{\rho}^{\textrm{s-}z}_{\textrm{c,NR}}$ commutes
with $H_{\textrm{SO}}$ and therefore served as a conserved quantity
to calculate the spin flux in the spin-orbit coupled system. With
the help of Eqs. (\ref{cs_nr_op}), (\ref{H_spin-orbital}), and the
Schr\"{o}dinger equation $\textrm{i}\hbar(\partial\psi/\partial t)
=H_{\textrm{SO}}\psi$, the continuity equation for the CSC in the NR
limit is obtained in the form
\begin{equation}\label{cs_nr_continue}
\frac{\partial \rho^{\textrm{s-}z}_{\textrm{c,NR}}}{\partial
t}+\bm{\nabla}\bullet\bm{j}^{\textrm{s-}z} _{\textrm{c,NR}}=0,
\end{equation}
with the current
\begin{equation}\label{cc_nr}
\bm{j}^{\textrm{s-}z} _{\textrm{c,NR}}=\frac{1}{4m}
\left[\psi^{\dag}\bm{\pi}\left(\hat{\rho}^{\textrm{s-}z}
_{\textrm{c,NR}}\psi\right)
+\left(\bm{\pi}\psi\right)^{\dag}\left(\hat{\rho}
^{\textrm{s-}z}_{\textrm{c,NR}}\psi\right)\right]+h.c. .
\end{equation}
Eqs. (\ref{cs_nr_op}), (\ref{cs_nr_continue}), and (\ref{cc_nr})
constitute the main results of this paper. Comparing to the
conventional spin current,\cite{Sinova} the CSC (\ref{cc_nr}) has an
additional correction term
\begin{equation}\label{j_corr}
\bm{j}_{\textrm{corr}}=\frac{\hbar}{16m^{3}c^{2}}
\left[\psi^{\dag}\bm{\pi}\hat{\mathcal{C}}\psi
+\left(\bm{\pi}\psi\right)^{\dag}\hat{\mathcal{C}}\psi+h.c.\right],
\end{equation}
which comes from the second term in Eq. (\ref{cs_nr_op}). One finds
that both the spin density and current have been modified to
construct the conserved CSC. The modification, the second term in
Eq. (\ref{cs_nr_op}) and Eq. (\ref{j_corr}), is a kind of the
relativistic correction resulting from the coupling between the
large and small components of the four-component spinor $\Psi$.
Those corrections are of the $(1/c^{2})$ order and are often small
enough to be neglected. However, the spin-orbit interaction is also
a relativistic correction of the $(1/c^{2})$ order. Therefore, the
CSC ought to be used instead of the bare spin current in the
spin-orbit coupled system.

Although the Hamiltonian (\ref{H_spin-orbital}) looks similar to
the Rashba model\cite{Rashba2,Bychkov} in the semiconductor
heterostructure, a realistic situation is quite complicated. The
Rashba coefficient $\alpha$ may not be directly determined by the
external electric field, but the asymmetry in the band structure
parameters of the heterostructure like the effective
mass.\cite{Zutic,Pfeffer} The discussion in this paper is
restricted to the case that the electron moves in a uniform space.
Therefore, it is interesting and desirable to apply the above idea
of the conserved CSC to the semiconductor heterostructures or
quantum-wires in future, though is highly non-trivial.

In summary, the CSC has been extracted from the TAM current by the
Gordon-like decomposition in the relativistic quantum mechanics
approach. The NR limit of the CSC is also derived. The CSC describes
the transport properties of the electron spin and is conserved in
the spin-orbit coupled system. It is also shown that the CSC remains
robust in the external electromagnetic field if both the electric
field and the magnetic field are along the same direction. As a
result, the CSC is a good substitution for the bare spin current in
studying the spin transport phenomena in spin-orbit coupled systems.

We thank J. C. Y. Teo and D. Y. Xing for valuable discussions.
This work was supported by the State Key Programs for Basic
Research of China, the National Natural Science Foundation of
China under Grant No. 10504011, the RGC funding of Hong Kong, and
Seed Funding for Basic Research of the University of Hong Kong.

\end{document}